\documentclass[10pt,twocolumn,floatfix,superscriptaddress,preprintnumbers,amsmath,amssymb,prb]{revtex4-2}
\usepackage{graphicx}% Include figure files
\usepackage{dcolumn}% Align table columns on decimal point
\usepackage{bm}% bold math
\usepackage{amssymb,amsmath}
\usepackage{float}
\usepackage{color}
\usepackage{ulem}

\begin{document}

\preprint{preprint(\today)}

\title{Probing the superconducting gap structure in the noncentrosymmetric topological superconductor ZrRuAs}

\author{Debarchan Das}
\email{debarchandas.phy@gmail.com}
\affiliation{Laboratory for Muon Spin Spectroscopy, Paul Scherrer Institute, CH-5232 Villigen PSI, Switzerland}

\author{D.T. Adroja}
\email{devashibhai.adroja@stfc.ac.uk}
\affiliation{ISIS Facility, Rutherford Appleton Laboratory, Chilton, Didcot, Oxon OX11 0QX, United Kingdom}
\affiliation{Highly Correlated Matter Research Group, Physics Department, University of Johannesburg, Auckland Park 2006, South Africa}
\author{M. R. Lees}
\affiliation{Department of Physics, University of Warwick, Coventry CV4 7AL, United Kingdom}
\author{R. W. Taylor}
\affiliation{Department of Physics, University of Warwick, Coventry CV4 7AL, United Kingdom}
\author{Z. S. Bishnoi}
\affiliation{Department of Physics, University of Warwick, Coventry CV4 7AL, United Kingdom}
\author{V. K. Anand}
\affiliation{Department of Physics, Indian Institute of Technology Delhi, Hauz Khas, New Delhi 110016, India}
\author{A. Bhattacharyya}
\affiliation{Department of Physics, Ramakrishna Mission Vivekananda Educational and Research Institute, Howrah 711202, India}
\author{Z. Guguchia}
\affiliation{Laboratory for Muon Spin Spectroscopy, Paul Scherrer Institute, CH-5232 Villigen PSI, Switzerland}
\author{C. Baines}
\affiliation{Laboratory for Muon Spin Spectroscopy, Paul Scherrer Institute, CH-5232 Villigen PSI, Switzerland}
\author{H. Luetkens}
\affiliation{Laboratory for Muon Spin Spectroscopy, Paul Scherrer Institute, CH-5232 Villigen PSI, Switzerland}
\author{G. B. G. Stenning}
\affiliation{ISIS Facility, Rutherford Appleton Laboratory, Chilton, Didcot, Oxon OX11 0QX, United Kingdom}
\author{Lei Duan}
\affiliation{The Institute of Physics, Chinese Academy of Science No. 8, Zhong Guan Cun South 3 Street, Haidian District, Beijing, postcode:100190, China}
\author{Xiancheng Wang}
\affiliation{The Institute of Physics, Chinese Academy of Science No. 8, Zhong Guan Cun South 3 Street, Haidian District, Beijing, postcode:100190, China}
\author{Changqing Jin}
\affiliation{The Institute of Physics, Chinese Academy of Science No. 8, Zhong Guan Cun South 3 Street, Haidian District, Beijing, postcode:100190, China}

\begin{abstract}

The superconducting gap structure of the topological superconductor candidate ZrRuAs with a noncentrosymmetric crystal structure has been investigated using muon spin rotation/relaxation ($\mu$SR) measurements in transverse-field (TF) and zero-field (ZF) geometries. Magnetization, electrical resistivity and heat capacity measurements reveal bulk superconductivity below a superconducting transition temperature $T_{\rm c} = 7.9(1)$~K. The temperature dependence of the effective penetration depth obtained from the TF-$\mu$SR spectra, and the electronic heat capacity in the superconducting state, are well described by an isotropic $s$-wave gap model. Comparison of the electronic mean free path with the superconducting coherence length suggests superconductivity in the dirty limit. ZF $\mu$SR data show there is no significant change in the muon spin relaxation rate above and below $T_{\rm c}$, indicating that time-reversal symmetry is preserved in the superconducting state.

\end{abstract}

%\pacs{74.20.Mn, 74.25.Ha, 74.70.Xa, 76.75.+i, 62.50.-p}

\maketitle

\section{Introduction}

The physics of topological superconductors (TSC) has attracted considerable attention~\cite{Qi,Ando, Tanaka, Sato, Hasan, Kitaev, Wilczek, Beenakker}. However, despite tremendous research activity, only a few materials with the potential to act as bulk topological superconductors, including Sr$_{2}$RuO$_{4}$~\cite{LukeTRS}, the Weyl semimetal $T_{d}$-MoTe$_{2}$~\cite{GuguchiaMoTe2}, Cu/Sr/Nb-doped Bi$_2$Se$_3$~\cite{Hor, Kriener, Smylie1, Kobayashi, Krieger, Das}, and PdTe$_2$~\cite{Noh}, have been identified.

In the search for new topological superconductors, the ternary transition metal pnictides $TT^{\prime}X$ ($T = $~Ca, Zr, Hf; $T^{\prime} = $~Ir, Ru, Ag and Os; $X =$~P, As and Si) offer considerable promise, as many members of this series exhibit interesting topological properties and some become superconducting~\cite{Barz1, Meisner1, Meisner2, Meisner3,Shirotani, Shirotani2, Ivanov, Yamakage}. The discovery of superconductivity with relatively high transition temperatures, $T_{\rm c}$, in ZrRuP ($T_{\rm c}$ = 13.0~K) and HfRuP ($T_{\rm c} = 12.7$~K)~\cite{Barz1, Meisner1} stimulated the study of related compounds, particularly with $T=$~Zr and Hf.

These $TT^{\prime}X$ compounds crystallize into one of four structure types~\cite{Johnson}: (i) a hexagonal Fe$_2$P-type $h$--phase, with space group $P\bar{6}2m$, (ii) an orthorhombic Co$_2$P-type $o$--phase, with space group $Pnma$, (iii) a hexagonal MgZn$_2$-type, with space group $P6_3/mmc$, or (iv) an orthorhombic TiFeSi-type $o^{\prime}$--phase, with space group $Ima2$. Even though superconductivity is found in both the $h$-- and $o$--phases, the $T_{\rm c}$'s, are generally higher for the $h$--phase. The Fe$_2$P $h$--phase lacks a centre of inversion and whether the noncentrosymmetric crystal structure plays an important role in governing the higher transition temperatures remains an open question. 

Two members of the Fe$_2$P-type are HfRuP ($T_{\rm c} = 12.7$~K) and ZrRuAs ($T_{\rm c}  = \operatorname{7.9-12}$~K)~\cite{Barz1,Meisner2, Qian}. Recent ARPES studies on these compounds reveal that HfRuP is a Weyl semimetal with 12 pairs of type--II Weyl points while ZrRuAs does not have any Weyl point and is suggested to belong to a topological crystalline insulating phase with nontrivial mirror Chern numbers~\cite{Qian}.

The noncentrosymmetric structure of these $h$--phase compounds containing heavy elements with strong spin-orbit interactions, also allows for the possibility of mixed spin-singlet and spin-triplet pairing~\cite{Bauer, Bauer1, Tanaka2}. Therefore, compounds crystallizing in the $h$--phase are of particular interest and an understanding of the superconducting gap structures of the $h$--phase members is essential. Recent studies on HfIrSi~\cite{Bhattacharyya} and ZrIrSi~\cite{Panda} that crystallize in the $o$-phase reveal $s$-wave superconductivity and preserved time-reversal symmetry. However, a detailed investigation of the pairing symmetry in the $h$-phase members including HfRuP or ZrRuAs is still lacking in the existing literature.

%%%%%%%%%%%%%%%%%%%%%%%%%%%%%%%%%%%%%%%%%%%%%%%%%%%%%%%%%%%%%%%
\begin{figure*}[htb!]
\includegraphics[width=1.0\linewidth]{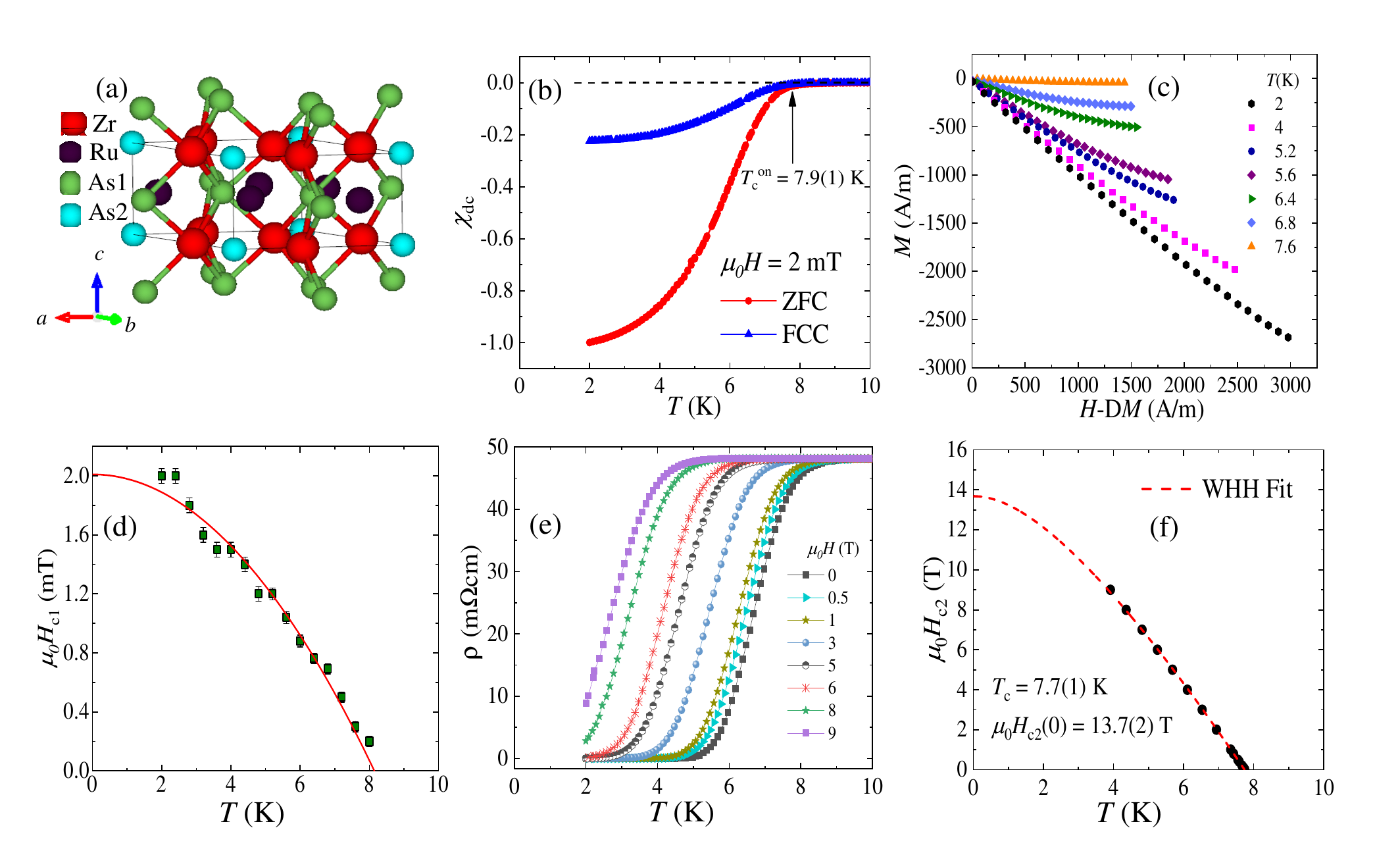}
%\vspace{-2.8cm}
\caption{(Color online) (a) Hexagonal crystal structure of ZrRuAs. Zr and Ru are represented by red and black spheres, respectively, and the two types of As environment are indicated by the green and cyan spheres. The unit cell is shown by the parallelepiped. (b) Temperature dependence of dc magnetic susceptibility collected in ZFC and FCC conditions. (c) Field dependence of the magnetization (corrected for demagnetization effects using $D=1/3$) at different fixed temperatures. (For clarity only selected curves are shown.) (d) Lower critical field, $H_{\rm c1}$, as a function of temperature. The red solid line represents a fit to the data with the Ginzburg-Landau expression as discussed in the text. (e) Electrical resistivity as a function of temperature at different applied fields. (For clarity, only selected fields data are shown.) (f) Upper critical field $H_{\rm c2}$ as a function of temperature derived from electrical transport measurements. The dashed line shows a fit with the WHH model, as discussed in the text.}
\label{fig1}
\end{figure*}
%%%%%%%%%%%%%%%%%%%%%%%%%%%%%%%%%%%%%%%%%%%%%%%%%%%%%%%%%%%%%%%%%%%%%%%%%%%%%%%%

In this report, we present the results of muon spin rotation/relaxation ($\mu$SR) measurements on ZrRuAs. $\mu$SR is a very sensitive technique to resolve the nature of the pairing in superconductors. For type-II superconductors, the mixed or vortex state gives rise to an inhomogeneous spatial distribution of local magnetic fields influencing the muon spin depolarization rate which is directly related to the magnetic penetration depth $\lambda$. Most importantly, the temperature dependence of $\lambda$ is particularly sensitive to the structure of the superconducting gap. Moreover, zero-field $\mu$SR is a very powerful tool for verifying whether or not time-reversal symmetry is preserved in the superconducting state. Here, the $\mu$SR results are supplemented by magnetization, $M$, electrical resistivity, $\rho$, and heat capacity, $C_{\rm p}$ measurements. Our results show $s$-wave pairing with preserved time-reversal symmetry in ZrRuAs.

\section{EXPERIMENTAL DETAILS }
A polycrystalline sample of ZrRuAs was prepared by a high pressure synthesis process. High purity ($>$99.999\%) lumps of As were ground into a fine powder and mixed with high purity ($>$99.99\%) powders of Zr and Ru in stoichiometric quantities. The resultant mixture was formed into a 6~mm diameter pellet, placed in a hexagonal boron nitride container, which was then packed in a pyrophyllite cube. The whole assembly was pressed to 5~GPa and heated to 1100~$^{\circ}$C, held at this temperature for half an hour. After that we quenched it to room temperature by switching off the power, while maintaining the pressure of 5 GPa, and cooling with circulating water.

Powder x-ray diffraction confirmed that the synthesized sample contains ZrRuAs as the main phase with small quantities of ZrAs ($\sim$~3\%) and unreacted Ru ($\sim$~5\%) as impurity phases. The x-ray pattern of the majority ZrRuAs phase was indexed to the hexagonal Fe$_2$P-type crystal structure with space group $P\bar{6}2m$ (No. 189). The measured lattice parameters are $a = 6.587(1)$~\AA~and $c = 3.886(1)$~\AA, in agreement with the published work~\cite{Meisner2, Qian}. It is to be noted that ZrAs has been reported to exhibit diamagnetic behavior with metallic resistivity without any sign of superconductivity down to 2~K~\cite{Saparov}.

Magnetization measurements were performed in a Quantum Design Magnetic Property Measurement System SQUID magnetometer under zero-field-cooled (ZFC) and field-cooled-cooling (FCC) conditions. Electrical transport and specific heat measurements were carried out in a Quantum~Design Physical Property Measurement System. Transverse-field (TF) and zero-field (ZF) ${\mu}$SR experiments were performed at the Paul Scherrer Institute (Villigen, Switzerland). The measurements down to 1.5~K were carried out on the GPS spectrometer and measurements down to 270~mK were performed on the DOLLY spectrometer. The sample was powdered and pressed into a 7~mm diameter pellet which was then mounted on a Cu holder using GE varnish. This holder assembly was then mounted in the appropriate spectrometer cryostat. Both spectrometers are equipped with a standard veto setup providing a low-background ${\mu}$SR signal. All the TF experiments were performed after field-cooled-cooling the sample. The ${\mu}$SR time spectra were analyzed using the MUSRFIT software package~\cite{Suter}.

\section{RESULTS AND DISCUSSION}

\subsection{Crystal Structure, Magnetization and Electrical Resistivity }
Figure~\ref{fig1}(a) shows the layered hexagonal structure of ZrRuAs which crystallizes in the $h$--phase with a space group $P\bar{6}2m$. Each layer in the hexagonal lattice is occupied by either Zr and As atoms or Ru and As atoms. Figure~\ref{fig1}(b) shows the temperature dependence of the ZFC and FCC dc magnetic susceptibility $\chi_{dc}(T)$ in an applied magnetic field of 2~mT. A clear diamagnetic signal is observed in both in ZFC and FCC curves with an onset superconducting critical temperature $T_{\rm c}^{\rm onset} = 7.9(1)$~K. It is to be noted that the $T_{\rm c}^{\rm onset}$ of our sample is very similar to that reported for the single crystal of ZrRuAs ($T_{\rm c}^{\rm onset} = 7.9$~K) by Qian \textit{et~al}.~\cite{Qian}, but lower than the $T_{\rm c}=12$~K reported by Meisner~\textit{et~al}.~\cite{Meisner2, Meisner3}, even though the crystal structure of the samples studied is the same. One possible reason for this discrepancy could be the different preparation methods used. It was argued by Qian~\textit{et~al}.~\cite{Qian} that the difference in annealing temperatures is responsible for the different $T_{\rm c}$'s. Following the same line of reasoning, it is also anticipated that the lower $T_{\rm c}$ in the sample is mainly due to the different heat treatment procedure. The sample magnetization, $M$, is presented as a function of applied field, $H$, in Fig.~\ref{fig1}(c) for temperatures between 2 and 7.6~K. The data have been corrected for demagnetization effects using a demagnetization factor $D= 1/3$ (appropriate for spherical powder grains). For applied fields below $H_{\rm{c1}}$, the sample is in the Meissner state. As the applied field strength is increased above $H_{\rm c1}$, the magnetic response deviates from linearity as the sample enters the vortex state, and flux lines start to penetrate the superconducting bulk. Values of $H_{\rm c1}(T)$ were determined as the fields at which $M$ deviates from linearity at a constant temperature, which were calculated following a procedure adapted from the method described in Ref.~[\onlinecite{Umezawa}]. The temperature dependence of the lower critical field [Fig.~\ref{fig1}(d)] can be modeled using the Ginzburg-Landau expression~\cite{Ginzburg} $H_{\rm c1}(T)= H_{\rm c1}(0)\left[1-(T/T_{\rm c})^2\right]$ yielding the value $\mu_0H_{\rm {c1}}(0)=2.01(5)$~mT and $T_{\rm c}^{H_{\rm c1}} = 8.1(1)$~K. The obtained value of $T_{\rm c}$ is very close to that determined from other measurements (see below).

%%%%%%%%%%%%%%%%%%%%%%%%%%%%%%%%%%%%%%%%%%%%%%%%%%%%%%%%%%%%%%%%%%%%%%%%%%%%%%%%%%%%%%%%%
\begin{figure}
\includegraphics[width=1.0\linewidth]{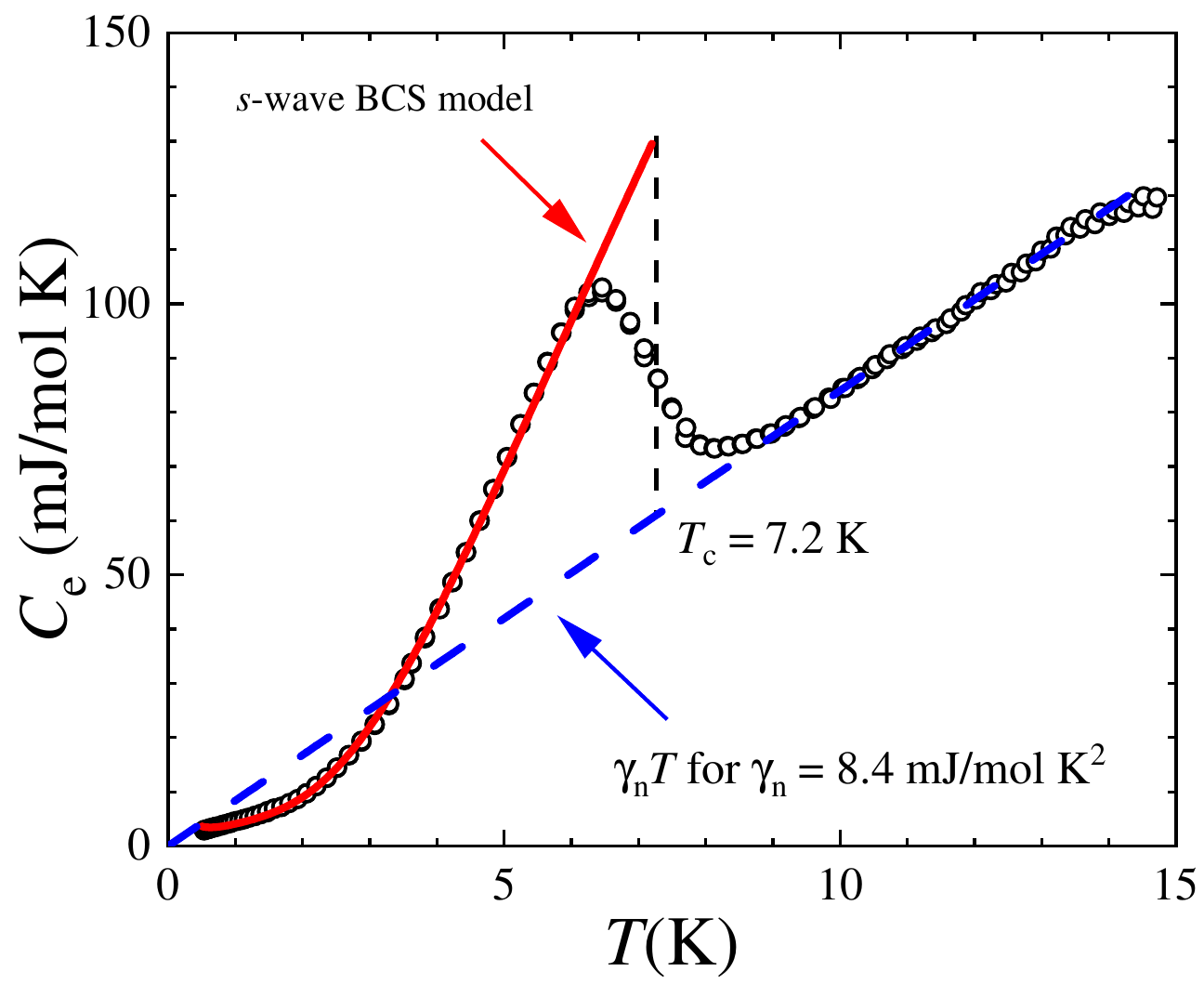}
\caption{(Color online) Electronic contribution $C_{\rm e}$ to the zero-field heat capacity of ZrRuAs as a function of temperature $T$. The solid red line is a fit using a Bardeen-Cooper-Schrieffer expression for a single-band $s$-wave superconductor including a contribution to $C_{\rm e}(T)$ from the portion of the sample that is non-superconducting.}
\label{fig:ZrRuAs_HC_el}
\end{figure}
%%%%%%%%%%%%%%%%%%%%%%%%%%%%%%%%%%%%%%%%%%%%%%%%%%%%%%%%%%%%%%%%%%%%%%%%%%%%%%%%%%%%%%%%%%%%%%%%

Figure~\ref{fig1}(e) presents the temperature dependence of the electrical resistivity $\rho(T)$ in the temperature range $0<T<10$~K under zero and different applied fields up to 9~T. The onset of the superconducting transition temperature matches very well with that obtained from magnetization measurements. The broad transition seen in the resistivity in zero field also agrees with the susceptibility results. Taking the temperature at 90\% of the residual resistivity at different applied fields gives the upper critical field-temperature ($\mu_0H_{\rm c2}(T)$) phase diagram for ZrRuAs presented in Fig.~\ref{fig1}(f). The temperature dependence of upper critical field can be modeled using the Werthamer-Helfand-Hohenberg (WHH) model~\cite{Werthamer} which accounts for Pauli limiting and spin-orbit scattering effects, that are expected to be strong in ZrRuAs as it is a non-centrosymmetric transition metal compound. The dashed line in Fig.~\ref{fig1}(f) shows a fit made with the WHH model yielding $\mu_0H_{\rm c2}(0)= 12.8(3)$~T. The Ginzburg-Landau coherence length, $\xi_{\rm GL}$, was calculated using the relation~\cite{Tinkham}  $\xi_{\rm GL}=\left[\Phi_0/(2\pi\mu_0 H_{\rm c2}(0)\right]^{1/2}$ where ${\Phi}_{\rm 0}=2.068 \times 10^{-15}$~Wb is the magnetic-flux quantum, giving $\xi_{\rm GL} = 5.08(3)$~nm. Using the value of effective penetration depth $\lambda$, estimated from the $\mu$SR measurements (see below), gives a Ginzburg-Landau parameter, $\kappa_{\rm{GL}}=\frac{\lambda}{\xi_{\rm GL}} \sim~94$ suggesting ZrRuAs is a strongly type II superconductor.

\subsection{Heat Capacity}
The bulk nature of the superconductivity in ZrRuAs is quite evident from the heat capacity data presented in Fig.~\ref{fig:ZrRuAs_HC_el} which shows the onset of superconductivity near $T_{\rm c}^{\rm on}$ = 7.90(5)~K. The normal state $C_{\rm p}(T)$ data are well described by $C_{\rm p}(T) = \gamma_{\rm n} T + \beta T^3 + \delta T^5$.  A fit of the zero-field $C_{\rm p}(T)$ data in the temperature range 10 to 15~K yields the normal state Sommerfeld coefficient $\gamma_{\rm n} = 8.4(2)$~mJ/mol\,K$^2$. The coefficient $\beta$ is found to  be 0.220(1)~mJ/mol\,K$^4$ which in turn provides an estimate of the Debye temperature $\Theta_{\rm D} = 298(1)$~K. The coefficient $\delta$ is found to be $1.52(9) \times 10^{-4}$~mJ/mol\,K$^6$~\cite{Gopal}.

The electronic contribution to the heat capacity $C_{\rm e}(T)$ after subtracting off the phonon contribution, $C_{\rm e}(T) = C_{\rm p}(T) - \beta T^3 - \delta T^5$ is shown in Fig.~\ref{fig:ZrRuAs_HC_el}. $C_{\rm e}(T)$ shows the bulk nature of the superconducting transition more clearly. The superconducting transition temperature corresponding to the entropy-conserving construction shown by the vertical dashed line in Fig.~\ref{fig:ZrRuAs_HC_el} is $T_{\rm c} = 7.2$~K. The $C_{\rm e}(T)$ data were fitted using a Bardeen-Cooper-Schrieffer (BCS) expression for a single-band $s$-wave superconductor~\cite{Tinkham, Poole2007}. The fit includes a contribution to $C_{\rm e}(T)$ from the portion of the sample that is non-superconducting. There is reasonable agreement between the experimental $C_{\rm e}(T)$ data and the BCS model as shown by the solid red line in Fig.~\ref{fig:ZrRuAs_HC_el}.

Using the value of $\Theta_{\rm D}= 298$~K, the electron-phonon coupling parameter $\lambda_{\rm e-ph}$ can be estimated from the McMillan theory~\cite{McMillan}
%%%%%%%%%%%%%%%%%%%%%%%%%%%%%%%%%%%%%%%%%%%%%%%%%%%%%%%%%%%%%%%
\begin{equation}
\lambda_{\rm e-ph}=\frac{1.04+\mu^*\ln(\Theta_{\rm D}/1.45T_{\rm c})}{(1-0.62\mu^*)\ln(\Theta_{\rm D}/1.45T_{\rm c})-1.04},
\label{eq1}
\end{equation}
%%%%%%%%%%%%%%%%%%%%%%%%%%%%%%%%%%%%%%%%%%%%%%%%%%%%%%%%%%%%%%%%
\noindent where $\mu^*$ is the Coulomb pseudopotential with a typical value of 0.13 for systems containing transition metals~\cite{Vora}. The value of $\lambda_{\rm e-ph}$  turns out to be 0.72(5). A similar value is also found in noncentrosymetric LaIrP ($\lambda_{\rm e-ph} = 0.67$), LaIrAs ($\lambda_{\rm e-ph} = 0.58$) and ThCoSi ($\lambda_{\rm e-ph} = 0.52$)~\cite{Qi2, Domieracki}. This low value of $\lambda_{\rm e-ph}$ is indicative of weak-coupling superconductivity in ZrRuAs.

\subsection{ TF-${\mu}$SR  Measurements}

%%%%%%%%%%%%%%%%%%%%%%%%%%%%%%%%%%%%%%%%%%%%%%%%%%%%%%%%%%%%%%%%%%%%%%%%%%%%%%%%%%%%%%%%
\begin{figure}[htb!]
\includegraphics[width=0.9\linewidth]{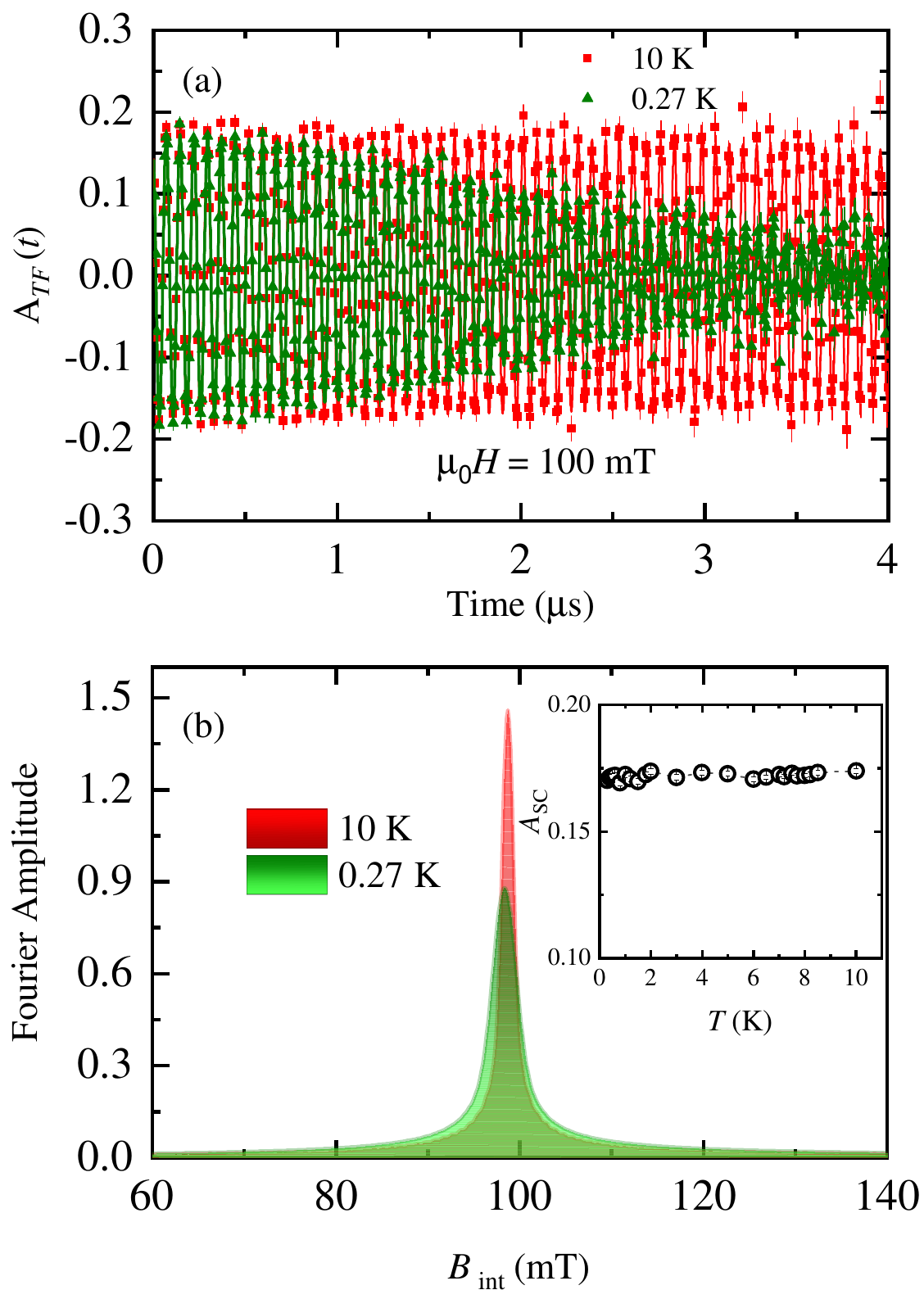}
\caption{(Color online)  (a) Transverse-field (TF) ${\mu}$SR time spectra obtained above and below $T_{\rm c}$ for {ZrRuAs} (after field cooling the sample from above $T_{\rm c}$). (b) Maximum entropy plots of the ${\mu}$SR time spectra from panel (a) at 0.27~K (green) and 10~K (red). The inset shows $A_{\rm SC}$ which is almost temperature independent between 0.27 and 10 K.}
\label{fig3}
\end{figure}
%%%%%%%%%%%%%%%%%%%%%%%%%%%%%%%%%%%%%%%%%%%%%%%%%%%%%%%%%%%%%%%%%%%%%%%%%%%%%%%%%%%%%%%%%%%%%%

Figure~\ref{fig3} compares the TF-$\mu$SR spectra for ZrRuAs at temperatures above (10~K) and below (0.27~K) $T_{\rm c}$, measured in an applied magnetic field of 100~mT. Above $T_{\rm c}$, TF-$\mu$SR spectra  show a small relaxation due to the presence of random local fields associated with the nuclear magnetic moments. In the superconducting state, the formation of the flux line lattice (FLL) causes an inhomogeneous distribution of magnetic field which increases the relaxation rate of the $\mu$SR signal. Assuming a Gaussian field distribution, observed TF-$\mu$SR asymmetry spectra can be analyzed using
%%%%%%%%%%%%%%%%%%%%%%%%%%%%%%%%%%%%%%%%%%%%%%%%%%%%%%%%%%%%%%%
%\begin{equation*}
\begin{multline}
A_{\rm TF}(t)=A_{\rm SC}\exp\left(\sigma^2t^2/2\right)\cos\left(\gamma_{\mu}B_{\rm int}t+\varphi\right) \\
+ A_{\rm NSC}\exp\left(-\Lambda_{\rm NSC}t\right)\cos\left(\gamma_{\mu}B_{\rm NSC}t+\varphi\right) ,
\label{ATF}
\end{multline}
%\end{equation*}
%%%%%%%%%%%%%%%%%%%%%%%%%%%%%%%%%%%%%%%%%%%%%%%%%%%%%%%%%%%%%%%%
\noindent where the first term  describes the oscillations (with a  Gaussian relaxation) produced by the superconducting (SC) fraction of the sample and the second term accounts for the non-superconducting (NSC) fraction (with a Lorentzian relaxation) from the impurities. $A_{\rm SC}$ and $A_{\rm NSC}$ denote the asymmetries related to SC and NSC fractions respectively. $\gamma_\mu/(2{\pi})\simeq 135.5$~MHz/T is the muon gyromagnetic ratio, and ${\varphi}$ is the initial phase of the muon-spin ensemble. $B_{\rm int}$ and $B_{\rm NSC}$ represent the internal magnetic fields at the muon site related to the SC and NSC fractions, respectively. The total relaxation rate $\sigma=\sqrt{\sigma_{\rm nm}^2+\sigma_{\rm SC}^2}$ where $\sigma_{\rm nm}$ and $\sigma_{\rm SC}$ represent the nuclear and superconducting vortex-lattice contributions, respectively. $\Lambda_{\rm NSC}$ is the relaxation rate related to the NSC fraction. The time domain spectra were fitted in two steps. First, the spectrum at 0.27~K was fitted using Eq.~\ref{ATF} in the time window 3 to $5~\mu$s giving $A_{\rm {NSC}}=0.012(4)$, $B_{\rm {NSC}}=103.6(3)$~mT (which is very close to the applied TF) and $\Lambda_{\rm NSC}=0.010(5)~\mu \rm{s}^{-1}$. In the second step, these parameters were fixed in Eq.~\ref{ATF} for subsequent fitting over the time window 0 to $5~\mu\rm{s}$ to obtain the temperature dependence of $\sigma$. Given the NSC impurities are non magnetic, this approach is justified. One should also note that $A_{\rm SC}$ remains almost temperature independent between 0.27 and 10 K [see inset of Fig.~3(b)]. The fits of the observed spectra with Eq.~\ref{ATF} are presented as solid lines in Fig.~\ref{fig3}(a). Figure~\ref{fig3}(b) shows the Fourier transform amplitudes of the TF-${\mu}$SR time spectra recorded at 10~K and 0.27~K. The sharp peak in the Fourier amplitude around 100~mT at 10~K corresponds to the external applied field. A fairly broad signal with a peak position slightly shifted to lower value (diamagnetic shift) shows that the sample is indeed in the superconducting mixed state. The formation of the FLL causes this broadening of the line shape.

%%%%%%%%%%%%%%%%%%%%%%%%%%%%%%%%%%%%%%%%%%%%%%%%%%%%%%%%%%%%%%%%%%%%%%%%%%%%%%%%
\begin{figure}[htb!]
\includegraphics[width=1.0\linewidth]{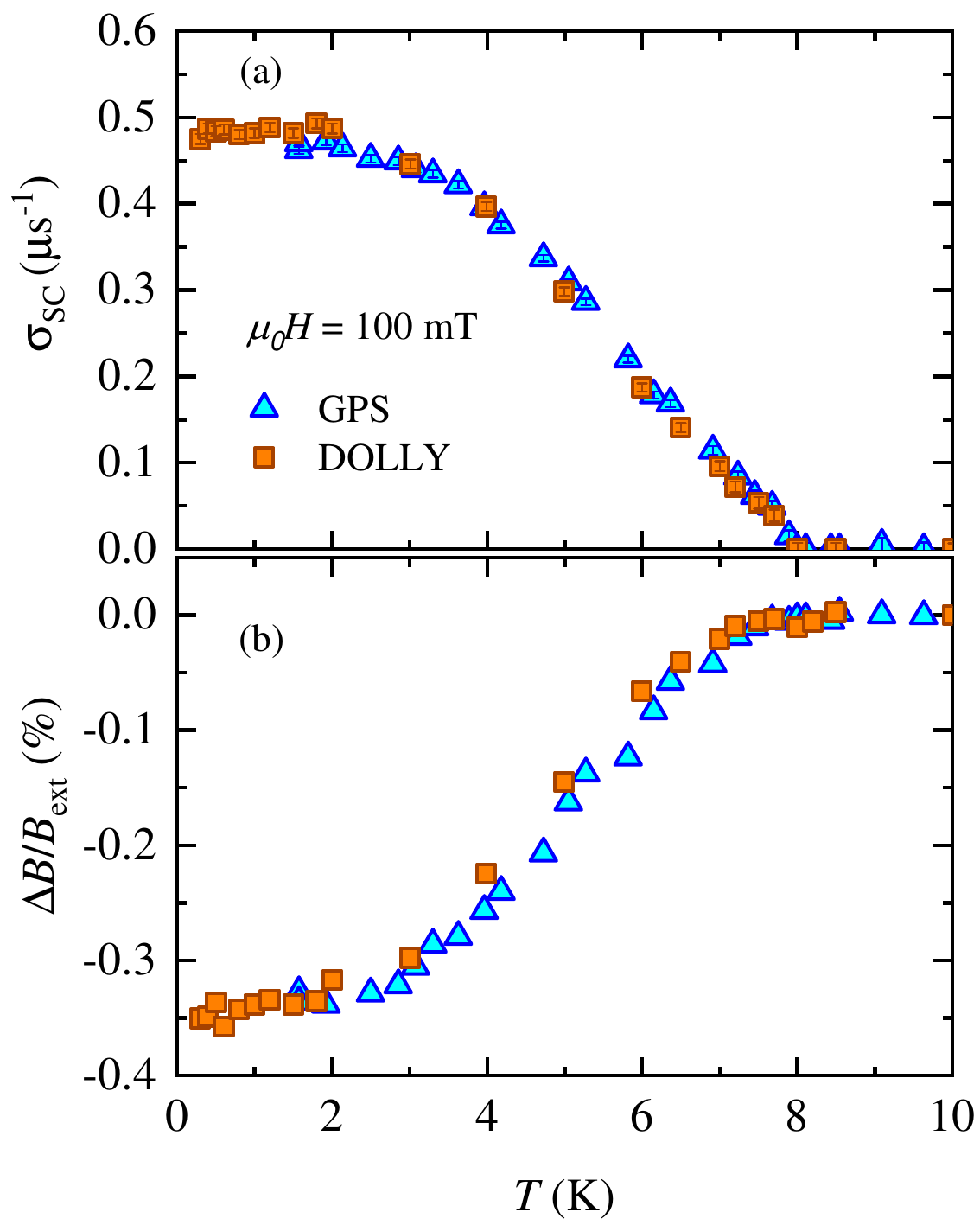}
\caption{(Color online) (a) Temperature dependence of the superconducting muon spin depolarization rate ${\sigma}_{\rm sc}$ of {\mbox ZrRuAs} measured in an applied magnetic field of ${\mu}_{\rm 0}H = 100$~mT. (b) Temperature dependence of the relative change of the internal field normalized to the external applied field, $\Delta B/B_{\rm ext}\left(= \frac {B_{\rm int}-B_{\rm ext}}{B_{\rm ext}}\right)$.}
\label{fig4}
\end{figure}
%%%%%%%%%%%%%%%%%%%%%%%%%%%%%%%%%%%%%%%%%%%%%%%%%%%%%%%%%%%%%%%

Figure~\ref{fig4}(a) shows the temperature dependence of ${\sigma}_{\rm sc}$ for ZrRuAs measured at an applied field of ${\mu}_{\rm 0}H=100$~mT. As seen from the figure, below $T_{\rm c}$, the relaxation rate ${\sigma}_{\rm sc}$ starts to increase from zero due to the formation of the FLL and saturates at low temperatures. The observed temperature dependence of ${\sigma}_{{\rm sc}}$, signals the presence of the single isotropic $s$-wave gap on the Fermi surface of ZrRuAs. Figure~\ref{fig4}(b) shows the temperature dependence of the relative change of the internal field normalized to the external applied field, $\Delta B/B_{\rm ext}\left(= \frac {B_{\rm \rm int}-B_{\rm ext}}{B_{\rm ext}}\right)$. It is quite evident that below $T_{\rm c}$, internal field values in the superconducting state are lower than the applied field due to a diamagnetic shift, as expected for type-II superconductors.

In the presence of a perfect triangular vortex lattice, the muon spin depolarization rate ${\sigma}_{\rm sc}(T)$ is directly related to the London magnetic penetration depth ${\lambda}(T)$ by~\cite{Brandt}:
%%%%%%%%%%%%%%%%%%%%%%%%%%%%%%%%%%%%%%%%%%%%%%%%%%%%%%%%%%%%%%%%%%%%%
\begin{equation}
\frac{\sigma_{\rm sc}^2(T)}{\gamma_\mu^2}=0.00371\frac{\Phi_0^2}{\lambda^4(T)}.
\label{Sigma}
\end{equation}
%%%%%%%%%%%%%%%%%%%%%%%%%%%%%%%%%%%%%%%%%%%%%%%%%%%%%%%%%%%%%%%%%%%%
\noindent Note, Eq.~\ref{Sigma} is only valid when the separation between the vortices is smaller than ${\lambda}$ which is presumed to be field independent in this model~\cite{Brandt}.

In order to reveal the superconducting gap structure of ZrRuAs and obtain quantitative estimates for the various parameters defining the superconducting state of this system, the temperature dependence of the magnetic penetration depth, ${\lambda}(T)$, which is directly associated with the superconducting gap was analyzed. As the temperature dependence of penetration depth shows clear saturation in the low-temperature regime, a $d$-wave scenario is excluded. 

Within the London approximation ($\lambda \gg {\xi}$), ${\lambda}(T)$ can be modeled for an $s$-wave superconductor in the clean limit using:~\cite{Suter,Tinkham, Prozorov}
%%%%%%%%%%%%%%%%%%%%%%%%%%%%%%%%%%%%%%%%%%%%%%%%%%%%%%%%%%%%%%%%%%%%%
\begin{equation}
\frac{\lambda^{-2}(T,\Delta_{0,i})}{\lambda^{-2}(0,\Delta_{0,i})}=
1+2\int_{\Delta(_{T})}^{\infty}\left(\frac{\partial f}{\partial E}\right)\frac{EdE}{\sqrt{E^2-\Delta_i(T)^2}},
\label{Lambda}
\end{equation}
%%%%%%%%%%%%%%%%%%%%%%%%%%%%%%%%%%%%%%%%%%%%%%%%%%%%%%%%%%%%%%%%%%%%
\noindent where $f=\left[1+\exp\left(E/k_{\rm B}T\right)\right]^{-1}$ is the Fermi function and ${\Delta}_{i}\left(T\right)={\Delta}_{0,i}{\Gamma}\left(T/T_{\rm c}\right)$. ${\Delta}_{0,i}$ is the maximum gap value at $T=0$~K.
The temperature dependence of the gap is described by the expression \mbox {${\Gamma}\left(T/T_{\rm c}\right)=\tanh\left\{1.82\left[1.018\left(T_{\rm c}/T-1\right)\right]^{0.51}\right\}$}~\cite{carrington}. 

In the case of an isotropic $s$-wave superconductor in the dirty limit~\cite{Tinkham}
%%%%%%%%%%%%%%%%%%%%%%%%%%%%%%%%%%%%%%%%%%%%%%%%%%%%%%%%%%%%%%%%%%%%%
\begin{equation}
\frac{\lambda^{-2}(T)}{\lambda^{-2}(0)}= \frac{\Delta (T)}{\Delta_0}\tanh\left[\frac{\Delta(T)}{2k_{\rm B}T}\right].
\label{BCS}
\end{equation}
%%%%%%%%%%%%%%%%%%%%%%%%%%%%%%%%%%%%%%%%%%%%%%%%%%%%%%%%%%%%%%%%%%%%

\begin{figure}[htb!]
\includegraphics[width=1.0\linewidth]{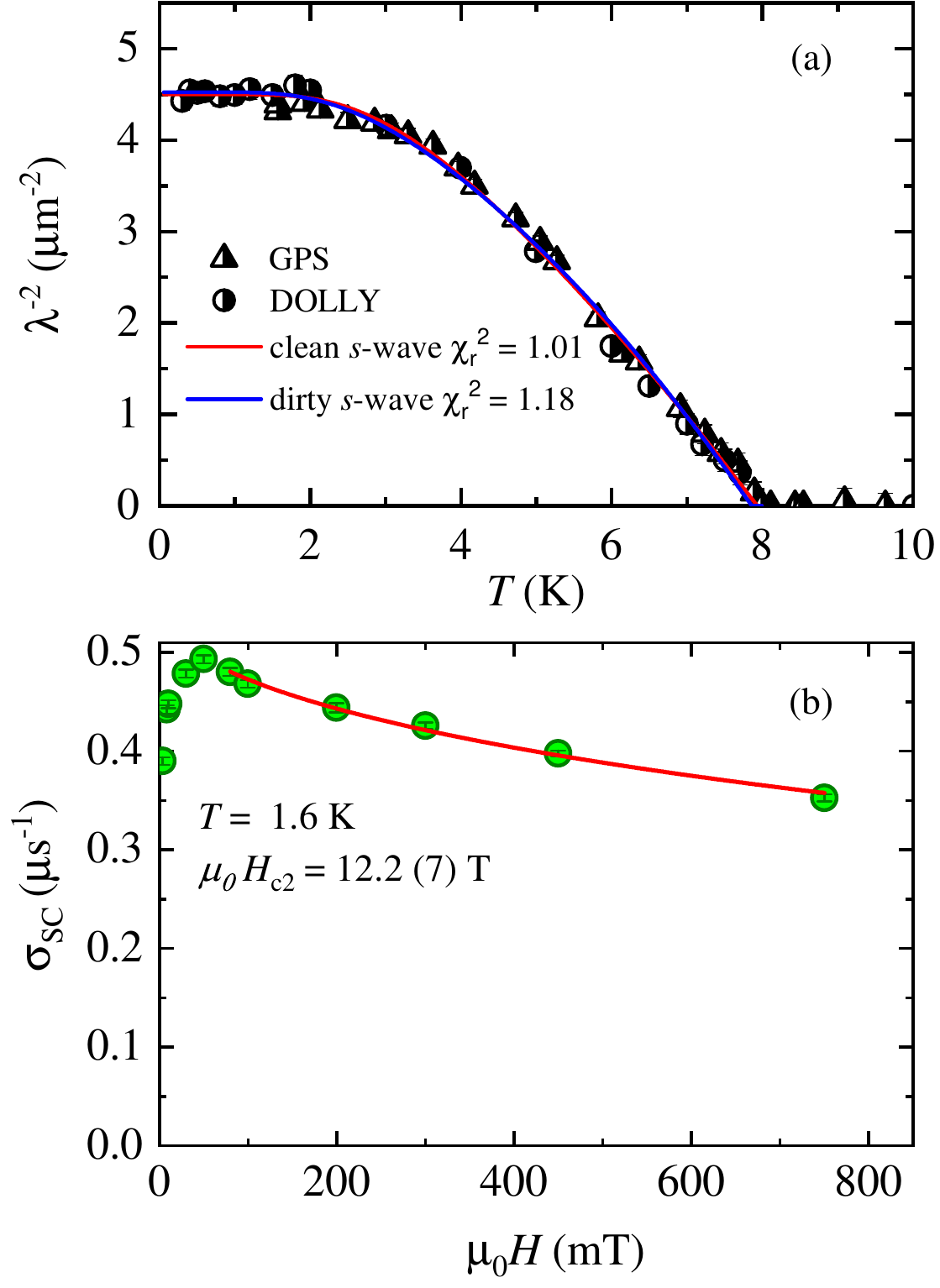}
\caption{(Color online) (a) Temperature dependence of ${\lambda}^{-2}$ for ZrRuAs, measured in an applied field ${\mu}_{\rm 0}H=100$~mT. The solid lines correspond to different theoretical models as discussed in the text. (b) Field dependence of the superconducting muon spin depolarization rate at 1.6~K fitted with an isotropic single $s$-wave gap model (solid line).}
\label{fig5}
\end{figure}
%%%%%%%%%%%%%%%%%%%%%%%%%%%%%%%%%%%%%%%%%%%%%%%%%%%%%%%%%%%%%%%

\begin{table*}[htb!]
	\centering
	\caption [Supercon para]{Superconducting parameters determined from fits to the temperature dependence of $\lambda(T)$ derived from TF-$\mu$SR experiments, using an $s$-wave model in clean and dirty limit.}
	\label{table:ZrRuAs-superconducting}
	\vskip .5cm
	\addtolength{\tabcolsep}{+5pt}
	\begin{tabular}{c c c c c c c c}
		\hline
		Model & $\Delta_0$ (meV) & $T_{\rm c}$ (K) & $\Delta_0/k_{\rm B}T_{\rm c}$ &$\lambda (0)$ (nm) & $n_{\rm s} (\times 10^{26}$m$^{-3}$)& $\chi^2_{\rm r}$ \\
		\hline
		Clean $s$-wave & 1.14(1) & 7.93(2) & 3.34(4) & 471(3) & 2.11(1) & 1.01\\
        Dirty $s$-wave & 0.87(2) & 7.89(3) & 2.56(7) & 470(2) & 2.20(2) & 1.18  \\
	%	$d$ &1.71(2) & 7.9 (1)& 442(2)&2.53(4)& 11.9 \\
		\hline
		\hline
	\end{tabular}
	\label{SCparameters}
\end{table*}

%(defined as $\chi^2_r = \chi^2$/(n-m); where n is the number of data points and m is the number of fit parameters) 

\noindent Figure~\ref{fig5}(a) shows the temperature evolution of ${\lambda}^{-2}(T)$ with the fits using the clean and dirty $s$-wave models. The superconducting gap parameters extracted from the fits are given in Table~\ref{SCparameters}. Both models (with different gap values) describe the observed temperature dependence of ${\lambda}^{-2}(T)$ reasonably well. The $s$-wave model in the clean limit gives a marginally lower value for the reduced $\chi^2_r$. To further address this point, the electronic mean free path, $\ell$, was estimated using the formalism given in Ref.~\cite{Barker}. Note, these calculations used the residual resistivity $\rho_0$ ($=0.16$~m$\Omega$~cm) of single crystal ZrRuAs~\cite{Qian} as the $\rho_0$ for the polycrystalline sample investigated here is high due to extrinsic factors such a poor intergranular connectivity. $\ell$ is estimated to be 3.1~nm which is comparable to $\xi_{\rm GL}$ indicating that the superconductivity in ZrRuAs is closer to the dirty limit. The gap value, $\Delta_0$, obtained for the dirty limit is slightly smaller than that for clean limit. The superconductivity in this compound appears to be fully gapped, consistent with the conclusions of electronic structure calculations~\cite{HM}.

Within the London theory~\cite{Sonier}, the penetration depth is directly related to microscopic quantities such as the effective mass, $m^*$, and the superconducting carrier density, $n_{\rm s}$ via the relation $\lambda^2(0)$=$\left( m^*/\mu_0n_{\rm s}e^2\right)$. $m^*$ can be estimated from the relation $m^*=(1+\lambda_{\rm e-ph})m_{\rm e}$ where $\lambda_{\rm e-ph}$ is the electron–phonon coupling constant and $m_{\rm e}$ is the electron rest mass. The values of $n_{\rm s}$ determined for the different models are also given in Table~\ref{SCparameters}. The values for $n_s$ are comparable to other $TT^{\prime}$X members such as HfIrSi ($n_s = 6.6 \times 10^{26}$~m$^{-3}$)~\cite{Bhattacharyya}, ZrIrSi ($n_s = 6.9\times 10^{26}$~m$^{-3}$)~\cite{Panda}. Interestingly, this value of $n_{\rm s}$ is also comparable to that seen in some other TSCs, e.g. Nb$_{0.25}$Bi$_2$Se$_3$  ($n_{\rm s} = 0.25\times 10^{26}~$m$^{-3}$)~\cite{Das} and $T_d$-MoTe$_2$ ($n_{\rm s} = 1.67\times 10^{26}~$m$^{-3}$)~\cite{GuguchiaMoTe2}. The relatively high value of $T_{\rm c}$ and low value of $n_{\rm s}$ in ZrRuAs signals possible unconventional superconductivity in this compound. It will be interesting to perform Hall conductivity measurements on this compound and compare the carrier density in the normal and superconducting states. This will be crucial to address the question of whether the single-gap superconductivity in ZrRuAs originates from the superconducting gap occurring only on an electron or hole-like Fermi surface. Furthermore, it is worth mentioning that for ZrRuAs, the ratio $T_{\rm c}\left[\rm K\right]/\lambda^{-2}(0) \left[\mu\rm{m}^{-2}\right] $ is 1.75 which is intermediate between the values observed for electron-doped ($T_{\rm c}/\lambda^{-2}(0)\sim~1$) and hole-doped ($T_{\rm c}/\lambda^{-2}(0)\sim~4$) cuprates \cite{Uemura,Uemura2, Shengelaya}. This may also be an indicative of unconventional superconductivity.
%\textcolor{green}{In order to probe the possible topological aspects of the superconducting properties in ZrRuAs, it will be important to investigate the depth dependent experiments on good quality single crystals using low energy muons.}

A fully gapped state is also evident from the field dependence of the TF-relaxation rate ${\sigma}_{\rm sc}(B)$. Figure~\ref{fig5}(b) shows ${\sigma}_{\rm sc}(B)$ at 1.6~K. Each point was obtained by field-cooling the sample from 10~K (above $T_{\rm c}$) to 1.6~K. As expected from the London model, initially, ${\sigma}_{\rm sc}$ rapidly increases with increasing magnetic field until reaching a maximum at 60~mT followed by a continuous decrease up to the highest field (750~mT) investigated. The field dependence follows the form expected for an $s$-wave superconductor with an ideal triangular vortex lattice. In addition, ${\sigma}_{\rm sc}(B)$ can also provide information about the upper critical field value. The observed ${\sigma}_{\rm sc}(B)$ curve at fields above the maximum, can be analyzed using the Brandt formula (for an $s$-wave superconductor)~\cite{Brandt2},
%%%%%%%%%%%%%%%%%%%%%%%%%%%%%%%%%%%%%%%%%%%%%%%%%%%%%%%%%%%%%%%%%%%%%

\begin{multline}
\sigma_{\mathrm{sc}}~\left[\mu \mathrm{s}^{-1}\right]= 4.83\times 10^4\left(1-\frac{H}{H_{\rm {c2}}}\right)\\
\times\left[1+1.21\left(1-\sqrt{\frac{H}{H_{\rm c2}}}\right)^3\right]\lambda^{-2}~\left[\rm {nm}^{-2}\right].
\label{Field_dep}
\end{multline}

%%%%%%%%%%%%%%%%%%%%%%%%%%%%%%%%%%%%%%%%%%%%%%%%%%%%%%%%%%%%%%%%%%%%
\noindent This gives an upper critical field of $\mu_0 H_{\rm c2}(0)=12.2(7)$~T which is in good agreement with the value obtained from the electrical resistivity measurements.

\subsection{ZF-${\mu}$SR Measurements}

%%%%%%%%%%%%%%%%%%%%%%%%%%%%%%%%%%%%%%%%%%%%%%%%%5
\begin{figure}[htb!]
\includegraphics[width=1.0\linewidth]{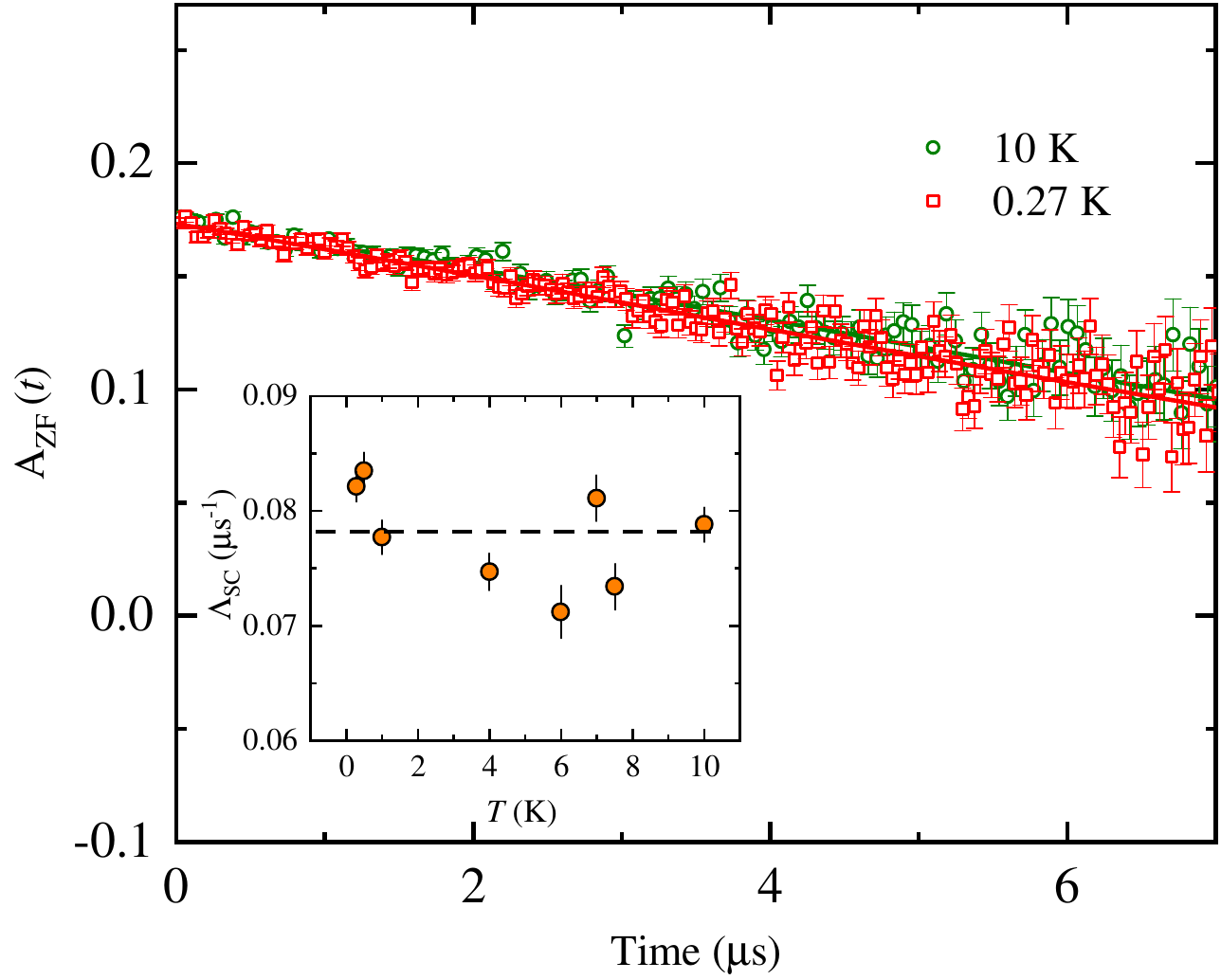}
\caption{(Color online) ZF $\mu$SR asymmetry spectra recorded at 0.27 and 10~K. Inset: Temperature dependence of the electronic relaxation rate measured in zero magnetic field for ZrRuAs.}
\label{fig6}
\end{figure}
%%%%%%%%%%%%%%%%%%%%%%%%%%%%%%%%%%%%%%%%%%%%%%%%%%%%%%%%%%%%%%%

ZF-${\mu}$SR experiments have also carried out above and below $T_{{\rm c}}$ to verify whether the time-reversal symmetry is preserved or not in ZrRuAs. The ZF-$\mu$SR spectra can be well described by
\begin{equation}
A_{\rm ZF}(t) = A_{\rm SC}\exp(-\Lambda_{\rm SC}t) + A_{\rm NSC}\exp(-\Lambda_{\rm NSC}t)
\label{ZF}
\end{equation}
\noindent where $A_{\rm NSC}$ and $\Lambda_{\rm NSC}$ were fixed to the values determined from the TF-measurements. Figure~\ref{fig6} shows that the $\mu$SR asymmetry spectra recorded above and below $T_{{\rm c}}$ show no noticeable change. The inset of Fig.~\ref{fig6} shows the temperature dependence of $\Lambda_{\rm SC}$ which shows no considerable enhancement across $T_{\rm c}$. Furthermore, the maximum possible spontaneous flux density due to superconductivity can be estimated using \mbox{($\Lambda_{\rm SC}|_{0.27~\rm K}-\Lambda_{\rm SC}|_{10~\rm K})/(2\pi\gamma_{\mu}) = 3.3~\mu$T} which is several times smaller than that seen for Sr$_2$RuO$_4$~\cite{LukeTRS}. Thus, time-reversal symmetry is most likely preserved in the superconducting state of ZrRuAs. This is consistent with $s$-wave superconductivity.

\section{SUMMARY}
Magnetic susceptibility, electrical resistivity, and heat capacity measurements on the topological superconductor candidate ZrRuAs reveal the onset of bulk type-II superconductivity at $T_{\rm c} = 7.9(1)$~K. $\mu_0H_{\rm c2}(T)$, determined from $\rho(T)$ and $C_{\rm p}(T)$ measurements made in various magnetic fields can be described by the WHH model. The temperature dependence of the magnetic penetration depth determined using TF-${\mu}$SR experiments is described very well by an $s$-wave model. The gap value, $2\Delta(0)$/$k_{\rm B}T_{\rm c}$, is slightly smaller than that expected for a BCS superconductor. ZF-$\mu$SR data reveal that time-reversal symmetry is preserved in the superconducting state of ZrRuAs.

%%%%%%%%%%%%%%%%%
\begin{acknowledgments}
The muon spectroscopy studies were performed at the Swiss Muon Source (S${\mu}$S) Paul Scherrer Insitute, Villigen, Switzerland. DTA would like to thank the Royal Society of London for Advanced Newton Fellowship funding between the UK and China, reference  NAF$\setminus$R1$\setminus$201248, and international exchange funding between UK and Japan. AB would like to acknowledge the Department of Science and Technology (DST) India for an Inspire Faculty Research Grant (DST/INSPIRE/04/2015/000169) and the SERB, India for core research grant support. The work in China is supported by the National Key R\&D Program of China and the National Natural Science Foundation of China.
\end{acknowledgments}

\end{document}